\documentclass[a4paper,11pt]{article}
\pdfoutput=1 

\usepackage{jcappub} 

\usepackage{enumerate}
\usepackage[normalem]{ulem}

\usepackage[T1]{fontenc} 
\usepackage{epstopdf}

\title{\boldmath Neutron stars in the braneworld within the Eddington-inspired Born-Infeld gravity}

\author[a,b]{I. Prasetyo}
\author[c]{I. Husin}
\author[a]{A. I. Qauli}
\author[a]{H. S. Ramadhan}
\author[a]{A. Sulaksono}

\affiliation[a]{Departemen Fisika, FMIPA, Universitas Indonesia, Depok 16424, Indonesia}
\affiliation[b]{Research Center for Physics, Indonesian Institute of Sciences (LIPI), Kompleks PUSPIPTEK Serpong, Tangerang 15310, Indonesia}
\affiliation[c]{Theoretical Physics Lab., THEPI Division, Institut Teknologi Bandung, Jl. Ganesha 10 Bandung 40132, Indonesia}

\emailAdd{ilham.prasetyo@sci.ui.ac.id}
\emailAdd{idrushusin@students.itb.ac.id}
\emailAdd{ali.ikhsanul@sci.ui.ac.id}
\emailAdd{hramad@ui.ac.id}
\emailAdd{anto.sulaksono@sci.ui.ac.id}

\abstract{
We propose the disappearance of ``the hyperon puzzle" in neutron star (NS) by invoking two new-physics prescriptions: modified gravity theory and braneworld scenario. By assuming that NS lives on a $3$-brane within a $5d$ empty AdS bulk, gravitationally governed by Eddington-inspired Born-Infeld (EiBI) theory, the field equations can be effectively cast into the usual Einstein's with ``apparent" anisotropic energy-momentum tensor. Solving the corresponding brane-TOV equations numerically, we study its mass-radius relation. It is known that the appearance of finite brane tension $\lambda$ reduces the compactness of the star. The compatibility of the braneworld results with observational constraints of NS mass and radius can be restored in our model by varying the EiBI's coupling constant, $\kappa$. We found that within the astrophysically-accepted range of parameters ($0<\kappa<6\times10^6\text{m}^2$ and $\lambda\gg1~ \text{MeV}^4$) the NS can have mass $\sim2.1~ \text{M}_\odot$ and radius $\sim10$ km.  
}

\begin{document}
\maketitle
\flushbottom

\def\Journal#1#2#3#4{{\it #1} {\bf #2}, #3 (#4) }
\def\RPP{{Rep. Prog. Phys}}
\def\PRC{{Phys. Rev. C}}
\def\PRD{{Phys. Rev. D}}
\def\ZPA{{Z. Phys. A}}
\def\NPA{{Nucl. Phys. A}} 
\def\JPG{{J. Phys. G }}
\def\PRL{{Phys. Rev. Lett}}
\def\PR{{Phys. Rep.}}
\def\PLB{{Phys. Lett. B}}
\def\AP{{Ann. Phys (N.Y.)}}
\def\EPJA{{Eur. Phys. J. A}}
\def\NP{{Nucl. Phys}}  
\def\RMP{{Rev. Mod. Phys}}
\def\IJMPE{{Int. J. Mod. Phys. E}}
\def\AJ{{Astrophys. J}}
\def\AJL{{Astrophys. J. Lett}}
\def\AA{{Astron. Astrophys}}
\def\ARAA{{Annu. Rev. Astron. Astrophys}}
\def\MPLA{{Mod. Phys. Lett. A}}
\def\ARNPS{{Annu. Rev. Nuc. Part. Sci}}
\def\LRR{{Living. Rev. Relativity}}
\def\CQG{{Class. Quantum. Grav}}

%
%
%
%

\section{Introduction}
\label{sec_intro}
Neutron stars (NSs) as the densest object in the universe are a unique avenue to study strong gravitational field and  matter under extreme conditions. Unfortunately, even though a lot of progress are reported,  until now the equation of state (EoS) of NSs are still quite poor to understand (see the discussions in Refs.~\cite{Lattimer2012,Chamel2013,Lonardoni,Yamamoto,Artyom14,ref:weissenborn,SB2012}). Furthermore, there is also a degeneracy between the EoS of NS matter and the theory of gravity applied to describe NS structure (see the discussions in Refs.~\cite{HFLN2015,Berti_etal2015} as well as the references therein). We note that among the modified theories of gravity, the Eddington-inspired Born-Infeld (EiBI) theory~\cite{Banados:2010ix} attracts attentions recently due to its particularity compared to general relativity (GR) (See Ref.~\cite{Berti_etal2015} for a review). One of the interesting features of EiBI theory is  the maximum mass value of NS can be adjusted by varying the corresponding $\kappa$ value.  Furthermore, through direct observations of the radii of low mass NS (around 0.5 $M_\odot$) and the measurements of neutron skin thickness of $^{208}$Pb, the EiBI theory can be discriminated from GR~\cite{Sotani14}. The range of  reasonable values of $\kappa$ parameter in EiBI theory can be constrained by using some astrophysical and cosmological data~\cite{Avelino12}, NSs properties~\cite{QISR2016,Harko13,PCD2011,PDC2012} and the Sun properties~\cite{CPLC2012}. Furthermore, in EiBI theory when matter can be described by a perfect fluid with barotropic EoS such as the one happens in compact stars, the modified field equations are equivalent to the one of GR.  In this view, the energy-momentum tensor can be still presented by an effective perfect fluid but with different behavior (It is called apparent EoS in Ref.~\cite{Delsate:2012ky}). It is found that even the EoS of matter in flat space time obeys all of the energy conditions, but the corresponding appeared EoS can violate some of energy conditions~\cite{Delsate:2012ky}. 

Observing the macroscopic properties of NS, and in particular their mass and radius can provide physical information about the composition of and interactions in NS matter including also the role of gravity inside NS. Recent analysis on the mass distribution of a number of pulsars with 
secure mass measurement has confirmed that $M \sim 2.1 M_\odot$ provides 
an established lower bound value on the maximum mass $(M_{\rm max})
$ of neutron star (NS) \cite{KKYT2013}.  We need to point out that our recent study~\cite{QISR2016} by using EiBI theory has shown that the $2.1 M_\odot$ maximum mass constraint  can be exceeded if we take value for $\kappa$  $\ge 4.0\times 10^{6}~\rm{m}^2$.  In that work, NS core EoS is calculated by using extended relativistic mean field model (ERMF) where standard SU(6) prescription and hyperons potential depths  are used to determine the hyperon coupling constants~\cite{QISR2016} while the crust EoS is taken from Ref.~\cite{MYN2013}.  On the other hand, the analysis methods used to extract the NS radii from observational data still suffer from  high uncertainty\cite{MCM2013,Bog2013, Gui2013,LS2013,Leahy2011,Steiner2010,Stein2013,Sule2011,Ozel2016}. Therefore, relative wide range of radius with mass around 1.4  $M_\odot$ has been reported (see the discussions in Ref.~\cite{JLF2015} and the references therein). From theoretical side, within GR the prediction of a canonical NS yields a radius between 10-15 km depending on the EoS model used (see details in Ref.~\cite{JLF2015,Miller2016} and the references therein). We need also to point out that the question that canonical NSs can have radii less than 10 km becomes a interesting topic of hot debate during last few years~\cite{JLF2015,Miller2016}. If such small NS radii are observed in the future, they will be very difficult to reconcile with existed EoS model estimates. Furthermore, it shown in Ref.~\cite{JLF2015} that within GR, only by introducing hypothetical particle such as a weak interacting light boson with appropriate in-medium parameters, the prediction of canonical NSs with radii less than 10 km can be realized. 

In cosmology, the proposal that universe is higher-dimensional is an old one, but it recently regains interest in the context of unification. The old Kaluza-Klein postulate was resurrected within the string-theory framework. A more phenomenological approach, known as the braneworld scenario, assumes that our four-dimensional spacetime is a slice of hyper-surface (called {\it the brane}) living in a higher-dimensional {\it bulk}~\cite{Rubakov:1983bb}. This set up was later proposed to solve the hierarchy problem, where the extra dimension needs not to be finite and periodic~\cite{Randall:1999ee, Randall:1999vf}. The description of NS in the braneworld within GR theory  is investigated previously in Refs.~\cite{Castro:2014xza,germani2001,OL2013,Bern2009}.  One of the interesting feature of this theory description is, the NS compactness in  this theory is less than that of standard GR. Therefore, for the same NS mass, the radius of NS in this theory is less than the one predicted by GR. It means that if in the future the NS with radius less than 10 km can be observed, it might be considered as the signature of  braneworld remnant.

In this work, we extend the previous works~\cite{Castro:2014xza,germani2001,OL2013,Bern2009}, by studying NS properties in the  braneworld within EiBI theory. Note that the nonlinearity of EiBI gravity makes it free from the ``hyperon puzzle"\footnote{Solutions to the hyperon puzzle have also been proposed in other framework of modified gravity. See, for example, Refs.~\cite{Astashenok:2014pua, Sakstein:2016oel}.}. We investigate also whether the mass and radius obtained are compatible to the known observational constraints as well as whether the apparent EoS within this theory obeys the energy conditions in~\cite{Delsate:2012ky}. The latter is quite relevant to investigate because the braneworld theory gives additional corrections in apparent pressure and apparent energy density of EiBI gravity. 

The paper is organized as follows. Sec.~\ref{sec_eibibraneworld}, is devoted to discussing  the field and TOV equations on the brane. Sec.~\ref{sec_numsol} discusses the numerical solutions and results obtained. The conclusion is given in Sec.~\ref{sec_conclu}.

\section{Brane with EiBI gravity in a bulk with GR}
\label{sec_eibibraneworld}
We consider a braneworld model~\cite{Shiromizu:1999wj} governed by the Eddington-inspired Born-Infeld (EiBI) theory. For readers who are not familiar with EiBI theory, see appendix~\ref{eibi}. Here we assume that the bulk is empty while (ordinary) matter lives on the brane, i.e., the energy momentum tensor in the brane has the modified form taken from EiBI theory.

We start from the {\it effective} Einstein equation for observer on the brane which has the following form~\cite{Shiromizu:1999wj,Maartens:2010ar}
\begin{equation}
\bar{G}_{\mu\nu} =8\pi G_N c^{-2} T^{\text{~eff}}_{\mu\nu}.
\end{equation}
The `bar' refers to quantities on the brane. The difference between the GR and EiBI gravity is that the Einstein tensor $\bar{G}^\mu_\nu$ is constructed from the auxiliary metric $q_{\mu\nu}$, not the physical one $g_{\mu\nu}$~\cite{Delsate:2012ky} (see appendix~\ref{eibi}). Meanwhile, the effective energy momentum tensor is
\begin{align}
T^{\text{~eff}}_{\mu\nu} =& -{\bar{\Lambda} c^{2} \over 8\pi G_N} q_{\mu\nu} + \bar{T}_{\mu\nu} + {\kappa_5^4c^2\over 8\pi G_N} \Pi_{\mu\nu} \nonumber\\
&-{c^{2}\over 8\pi G_N}\mathcal{E}_{\mu\nu}+{\kappa^2_5\over 8\pi G_N c^{-2}} W_{\mu\nu}.\label{eq:Teff}
\end{align}
This formula comes from projecting gravity in a 5-dimensional bulk spacetime into the 4-dimensional brane. Respectively, each has gravitational constant $\kappa^2_5$ and $G_N$ while the cosmological constants are $\Lambda$ and $\bar{\Lambda}$. Here $c$ is the speed of light and $\lambda$ is the tension of the brane observed in the bulk. 
The cosmological constant and gravitational constant from the brane are related to the ones from the bulk by
\begin{equation}
\bar{\Lambda}=\kappa^2_5\left({\Lambda\over 2} + {\lambda^2 \over 12}\right),~~G_N={\kappa^4_5 c^2\lambda \over 48\pi}.
\end{equation}
The second term on right side of Eq.~\eqref{eq:Teff}, $\bar{T}_{\mu\nu}$, is the energy-momentum tensor on the brane, modified by the EiBI gravity with nonlinearity constant $\kappa$ (see appendix \ref{eibi}), whose form is
\begin{align}
\bar{T}_{\mu\nu}= \tau T^\kappa_\mu q_{\kappa\nu}-\left(\tau {T\over2}+{c^2\over 8\pi G_N}{(1-\tau\eta)\over \kappa} \right) q_{\mu\nu}, \label{eq:apparentT}
\end{align}
with $T^{\mu}_{\nu}\equiv T_{\kappa\nu}g^{\kappa\mu}$,
\begin{equation}
\tau\equiv{1\over \sqrt{\det\left(\eta \delta^\mu_\nu-{8\pi G_N\over c^2}\kappa T^\mu_\nu\right)}},
\end{equation}
and
\begin{equation}
T_{\mu\nu}=(\rho+pc^{-2}) v_\mu v_\nu + pc^{-2} g_{\mu\nu},
\end{equation}
whose velocity vector is $v^a=(\sqrt{-g^{00}},0,0,0)$ and $v^a v_a=-1$. Notice that $T_{\mu\nu}$ is coupled to $g_{\mu\nu}$, and this metric is related to $q_{\mu\nu}$ by \eqref{eq:relationmetric} in the appendix \ref{eibi}. At the limit $\kappa\rightarrow0$, $\bar{T}_{mn}\rightarrow T_{mn}$.
$\Pi_{mn}$ is the local correction term, whose form is
\begin{equation}
\Pi_{\mu\nu}= {-6\bar{T}^a_{\mu}\bar{T}_{a\nu} +2\bar{T}\bar{T}_{\mu\nu}+ (3\bar{T}^{\alpha\beta}\bar{T}_{\alpha\beta} -\bar{T}^2)q_{\mu\nu} \over 24}.
\end{equation}
$\mathcal{E}_{mn}$ and $W_{mn}$ are, respectively, the nonlocal correction from the bulk's geometry and matter. The former is the bulk Weyl tensor contribution
\begin{equation}
\mathcal{E}_{\mu\nu}=C_{abcd} n^a n^c q^b_\mu q^d_\nu,
\end{equation}
while the latter is the contribution from matter in the bulk (whose energy-momentum tensor is $\hat{T}_{ab}$)
\begin{equation}
W_{\mu\nu}={2\over 3} \left(q^a_\mu q^b_\nu + n^a n^b q_{\mu\nu} - {q_{\mu\nu} \over 4}h^{ab}\right) \hat{T}_{ab},
\end{equation}
with $h_{ab}$ is the metric from the 5-dimensional bulk defined as
\begin{equation}
h_{ab}dx^adx^b=q_{\mu\nu}dx^\mu dx^\nu+d\chi^2
\end{equation}
with the brane is positioned at $\chi=0$.
Notice that as $\kappa_5\rightarrow 0$ we want this equation to become the Einstein's equation, so the second term must still be there. Thus, the brane has very high positive tension, $\lambda \rightarrow \infty$.

In this paper we assume that there is no matter in the bulk (implying $W_{mn}=0$) and the brane's cosmological constant is set to zero $\bar{\Lambda}=0$, which makes the tension of the brane become $\lambda=48\pi G_N/\kappa^4_5 c^2$ and ${\Lambda} =- {2^{7}\cdot 3\pi^2 G_N^2/\kappa_5^8 c^4}$ hence the bulk's geometry must be anti-de Sitter. [This set-up is reminiscent of the Randall-Sundrum fine-tuning problem~\cite{Randall:1999vf}. We shall have something to say on this in the conclusions.] These then simplify the effective energy-momentum tensor to be
\begin{equation}
{T^\mu_\nu}_\text{eff}= \bar{T}^\mu_\nu + {6\over \lambda} \Pi^\mu_\nu -{1\over 8\pi G_N c^{-2}}\mathcal{E}^\mu_\nu,
\end{equation}
The nonlocal correction is defined to have static spherical symmetry from the so-called ``Weyl fluid''~\cite{germani2001,maartens2000,Maartens:2010ar}
\begin{equation}
\mathcal{E}^\mu_\nu=-{6\left[
\mathcal{U}v^\mu v_\nu + \mathcal{P} r^\mu r_\nu + {(\mathcal{U}-\mathcal{P})\over 3} (\delta^\mu_\nu+v^\mu v_\nu)
\right]\over 8\pi G_N c^{-2} \lambda},
\end{equation}
where $\mathcal{U}$  and $\mathcal{P}$ are the nonlocal energy density and the anisotropic stress seen by observer in the brane, respectively, while $r^m$ is a unit radial 3-vector ($r^r r_r=1$, else $=0$).

Explicitly, the components of the effective energy-momentum tensor are
\begin{eqnarray}
{T^0_0}_\text{eff}&=& -\rho_\text{eff},\label{eq:momten1}\\
{T^r_r}_\text{eff}&=& p_\text{eff} c^{-2}+{4\mathcal{P}\over (8\pi G_N c^{-2})^2 \lambda},\\
{T^\theta_\theta}_\text{eff}&=& p_\text{eff} c^{-2}-{2\mathcal{P}\over (8\pi G_N c^{-2})^2 \lambda}.\label{eq:momten3}
\end{eqnarray}
with effective energy density and pressure are, respectively, defined as
\begin{eqnarray}
\rho_\text{eff}&=& \rho_q + {\rho_q^2\over 2\lambda} + {6\mathcal{U}\over (8\pi G_N c^{-2})^2 \lambda},\label{eq:edeneff}\\
p_\text{eff} c^{-2}&=& p_q c^{-2} + {\rho_q(\rho_q+2p_q c^{-2})\over 2\lambda} + {2\mathcal{U}\over (8\pi G_N c^{-2})^2 \lambda},\label{eq:presseff}
\end{eqnarray}
with $\rho_q\equiv-\bar{T}^0_0$ and $p_q\equiv\bar{T}^r_r$.
The metric is defined in~\eqref{eq:metricq} to also have spherical symmetry. 
So the Ricci tensor components are
\begin{eqnarray}
R^0_0 &=& -G^{-2}\left[{F''\over F}-{F'G'\over FG} +2{F'\over Fr}\right],\label{eq:BE01}\\
R^r_r &=& -G^{-2}\left[{F''\over F}-{F'G'\over FG} -2 {G'\over Gr}\right],\label{eq:BE02}\\
R^\theta_\theta&=& -G^{-2}\left[
{F'\over Fr}-{G'\over Gr}+{1\over r^2} \right]+{1\over r^2},\label{eq:BE03}
\end{eqnarray}
The Einstein's equations then has the form
\begin{eqnarray}
R^0_0&=&8\pi G_N c^{-2} \left({{T^0_0}_\text{eff}-{T^r_r}_\text{eff}-2{T^\theta_\theta}_\text{eff} \over2} \right),\\
R^r_r&=&8\pi G_N c^{-2} \left({-{T^0_0}_\text{eff}+{T^r_r}_\text{eff}-2{T^\theta_\theta}_\text{eff} \over2} \right),\\
R^\theta_\theta&=&8\pi G_N c^{-2} \left({-{T^0_0}_\text{eff}-{T^r_r}_\text{eff} \over2} \right).
\end{eqnarray}
Defining the metric solution $G^{-2}=1-{2G_N c^{-2}m(r)\over r}$ to be dependent on a mass function
\begin{equation}
m'(r)=4\pi {r}^2  \rho_\text{eff}.\label{eq:0001}
\end{equation}
the Einstein's equations imply
\begin{equation}
{F'\over F}={r^3(8\pi G_N c^{-2}) \left[p_\text{eff}c^{-2}+{4\mathcal{P}\over (8\pi G_N c^{-2})^2 \lambda} \right]+2G_N c^{-2}m \over 2r(r-2G_N c^{-2}m)}.
\label{eq:0002}
\end{equation}
The Bianchi identity of the brane $\nabla_m \bar{T}^{mn}=0$ and the bulk $\nabla_m {T}^{mn}_\text{eff}=0$, respectively, imply
\begin{align}
p'(r)
=&-{F'\over F}{b\over 2\pi G_N c^{-4} \kappa} \nonumber\\
&\times\left[
{ab(a^2-b^2) \over {4ab^2+(3a-bc_q^2)(a^2-b^2)}}  
\right],\label{eq:0003}
\end{align}
with $c_q^2={da\over db}=-{b\over a}{d\rho\over d(pc^{-2})}$ is the speed of sound and 
\begin{eqnarray}
\left[{\mathcal{U}+2\mathcal{P}} \right]'
&=& -{F'\over F} \left[ {4\mathcal{U}+2\mathcal{P}} \right] -{6\mathcal{P}\over r}\nonumber\\
&&- {(8\pi G_N c^{-2})^2\over 2} \left[ {\rho_q+p_q c^{-2} }\right] \rho_q'. 
\label{eq:eq04}
\end{eqnarray}
There are six solutions but only four equations of motion, thus we need to have two constraints (or equations of state): $p=p(\rho)$ 
and $\mathcal{P}=\mathcal{P}(\mathcal{U})$. Here we consider the first one from neutron star matter and the second one is modeled by $\mathcal{P}=w\mathcal{U}$ with $w$ a constant (as also used in~\cite{Castro:2014xza,Felipe:2016lvp}) which makes Eq. \eqref{eq:eq04} becomes
\begin{eqnarray}
\mathcal{U}'
= {-2\over \left[{1+2w } \right]}&&\left({F'\over F} \left[ {2+w} \right]\mathcal{U} +{3w\over r}\mathcal{U}\right.\nonumber\\
&&+\left.(4\pi G_N c^{-2})^2 \left[ {\rho_q+p_q c^{-2} }\right] \rho_q'\right).~~
\label{eq:0004}
\end{eqnarray}

\section{Numerical solutions}
\label{sec_numsol}

Here we calculate numerically the TOV equations for the physical mass (Eq.\eqref{eq:0001}), the physical pressure (Eq. \eqref{eq:0003}), the Weyl energy $\mathcal{U}$ (Eq.\eqref{eq:0004}), and the metric profile $F$ (Eq.\eqref{eq:0002}), which are expressed by  first-order ordinary differential equations, using Runge-Kutta fourth-order algorithm. We use the EoS $\epsilon=\rho c^{-2}=\epsilon(p)$ for NS matter based on BSR23 parameter set of ERMF model where the standard SU(6) prescription and hyperons potential depths  are utilized to determine the hyperon coupling constants, and the crust EoS is taken from Ref.~\cite{MYN2013} (see the details of NS EoS model for example in Ref.~\cite{QISR2016}). We impose the following boundary conditions:
\begin{equation}
p(r\rightarrow0)=p_c,~~ m(r\rightarrow0)=0,~~ \mathcal{\tilde{U}}(r\rightarrow0)=0.
\end{equation}
The numerical calculation starts from $r\sim0$ with arbitrary $p_c$ to the surface of the star at $r=R$ when $p(R)\sim0$. Note that here the NS mass has dimension of solar mass unit $M_\odot$, pressure $p$ and energy density $\epsilon=\rho c^{2}$ have dimension MeV/fm$^{3}$, and both Weyl energy $\mathcal{U}$ and Weyl anisotropic stress $\mathcal{P}$ have dimension m$^{-4}$ due to using natural units. We redefine the Weyl energy and Weyl anisotropic stress, respectively, by $\mathrm{U}\equiv\mathcal{U}/(8\pi G_Nc^{-2})^2$ and $\mathrm{P}\equiv\mathcal{P}/(8\pi G_Nc^{-2})^2$ to make things simpler. Both $\mathrm{U}$ and $\mathrm{P}$ have the same unit MeV$^2$/fm$^{6}$. The EiBI parameter $\kappa$ and the braneworld parameters $\lambda$ has dimension $10^6$ m$^2$ and MeV/fm$^{3}$, respectively, while $w$ is dimensionless and here we take it to be in a range of $-3\leq w\leq2$. We need to note that the authors of Ref.~\cite{Castro:2014xza} have found that $\lambda$ clearly controls the value of NS maximum masses, while  $w$ influences the corresponding radii. They have established a range of  $\lambda$ between 0.4 $\times 10^{37}$ dyne/cm$^2$ and 10 $\times 10^{37}$ dyne/cm$^2$, where $10^{37}$ dyne/cm$^2$ $=6.24\times10^3$ MeV/fm$^3$. This value is higher than the lower bound obtained from the gravitational wave calculation~\cite{Garcia-Aspeitia:2013jea}. In EiBI theory the parameter  $\kappa$ is also controls  the value of NS maximum masses~\cite{QISR2016}. Therefore, in the braneworld within EiBI theory, there are two parameters which control the value of NS maximum masses. It is common to ignore the EiBI's cosmological constant $H=(\eta-1)/\kappa$ by setting $\eta=1$ when discussing compact objects. In this work we have relaxed this assumption by considering also $\eta<1$ and $\eta>1$ cases which imply de Sitter and anti-de Sitter background in the brane, respectively, and investigate their impact on the NS properties. 

Note that we denote the physical mass as $M(r)$, which is related to the auxiliary mass $m(r)$ in \eqref{eq:0001}by \eqref{eq:relation01}, i.e.
\begin{equation}
B^2=G^2/ab,
\end{equation}
with $a$ and $b$ given in \eqref{eq:ab}.
Since it is usual to define the metric solution to have the form
\begin{equation}
B^{-2}=1-{2G_Nc^{-2}M(r)\over r},
\end{equation}
we have
\begin{equation}
M(r)={1-ab\over 2G_N c^{-2}}r+ab m(r).
\end{equation}
In the case of mass-radius relation graphs, the mass is specified at the surface $M(R)$ where $a$ and $b$ both goes to the constant $\eta^{-1/2}$ hence we can use
\begin{equation}
M(R)={1-\eta\over 2G_N c^{-2}}R+\eta m(R).
\end{equation}
This implies $M(R)=m(r)$ at the limit $\eta\to1$.

\begin{figure}[t]
\centering
\includegraphics[width=.45\linewidth]{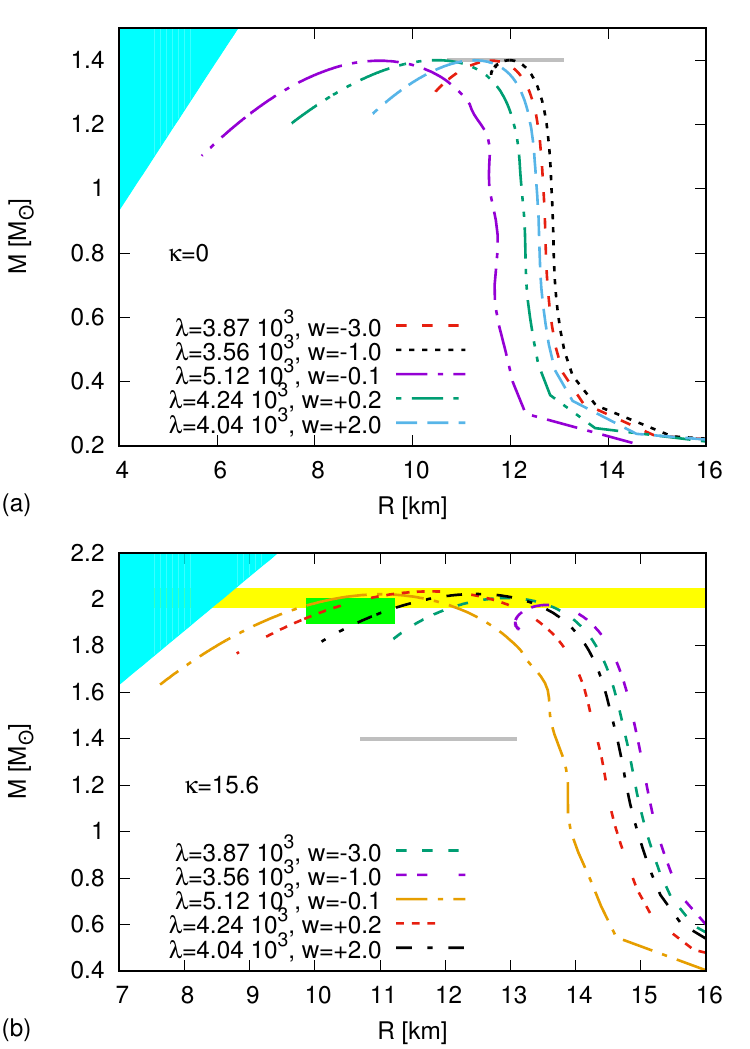}
\caption{Mass-radius relation without EiBI parameter $\kappa\rightarrow0$ (a) and with EiBI parameter $\kappa=15.6$ (b) for the values of $w,\lambda$ varied. (Here we use $\eta=1$.) The gray, yellow, and green shaded regions  are from observational constraints of Ref. \cite{Steiner:2010fz} for radius constraint, \cite{Antoniadis:2013pzd} for maximum mass constraint,and \cite{Ozel:2016oaf} for constraint obtained from simultan analysis of masses and radii of some pulsars, while the cyan one is the causality region from Ref. \cite{Lattimer:2006xb}.}
\label{fig:k=1d0 and k=15.6d6}
\end{figure} 

\begin{figure}[t]
\centering
\includegraphics[width=.45\linewidth]{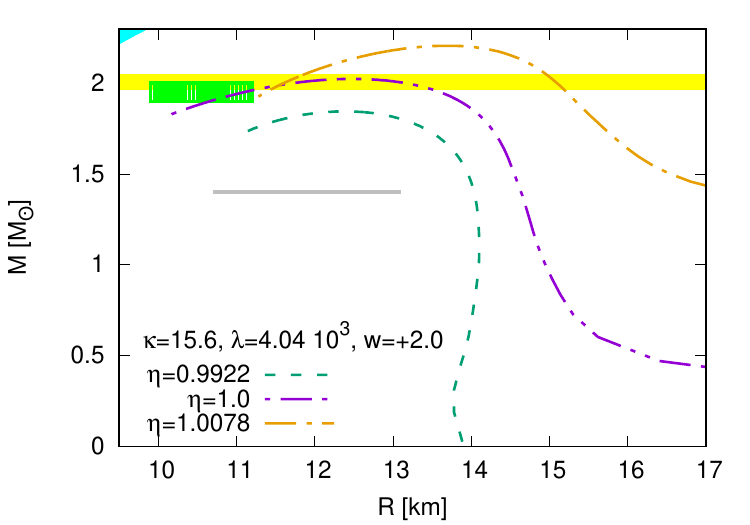}
\caption{Mass-radius relation with $\eta$ varied.}
\label{fig:eta_neq_1}
\end{figure}

In Fig.~\ref{fig:k=1d0 and k=15.6d6}, we show the mass-radius relation predicted by this model with EiBI parameter set to be $\kappa\rightarrow0$ which is equivalent to braneworld within GR and the one with EiBI parameter $\kappa=15.6$ for the values of $w,\lambda$ are varied. Here we use $\eta=1$. Respectively, the gray, yellow, and green shaded regions are the observational constraints from  Ref. \cite{Steiner:2010fz} for radius constraint, 	\cite{Antoniadis:2013pzd} for maximum mass constraint, and \cite{Ozel:2016oaf} for constraint obtained from simultaneous analysis of masses and radii of some pulsars, while the cyan one is the causality region from Ref. \cite{Lattimer:2006xb}. In this theory, if one increases $\lambda$ and $\kappa$ then the mass also increases. However, one can adjust $w$ value to vary the radius while keeping the mass $m(R) \simeq 1.4 M_\odot$. This can be seen clearly in (a) of Fig.~\ref{fig:k=1d0 and k=15.6d6} when $\kappa\rightarrow0$. The same case while  keeping the mass $m(R) \simeq 2.1 M_\odot$ in (b) of Fig. \ref{fig:k=1d0 and k=15.6d6} when $\kappa=15.6$. It can also be observed in (a) of Fig.~\ref{fig:k=1d0 and k=15.6d6} that at some $w$ values the radius is compatible with the observed region constraint from Ref. \cite{Steiner:2010fz} and there is also present canonical NS with radius $\le$ 10 km. To increase the NS maximum mass to be near $2.1 M_\odot$, we can increase $\kappa$ value as shown in (b) of Fig. \ref{fig:k=1d0 and k=15.6d6}. This has the maximum mass and radius of the star in one of the regions of observational constraints of Ref. \cite{Antoniadis:2013pzd} and  Ref. \cite{Ozel:2016oaf}. However, within the used parameters of the corresponding model, it is still difficult to obey simultaneously all three masses and radii constraints of Refs. \cite{Steiner:2010fz,Ozel:2016oaf,Antoniadis:2013pzd} by varying  $w$. The mass-radius relation obtained by varying $\eta$ are shown in Fig. \ref{fig:eta_neq_1}. By slightly increasing (decreasing) $\eta$,  the NS maximum mass is significantly increases (decreases). It seems that small radii NS prefer $\eta<1$ more that $\eta>1$.
{This happens when $\kappa$ is huge, whose implication is the brane has anti-de Sitter background with strong EiBI gravity nonlinearity with much bulk contribution. This is not always the case since we can also obtain the maximum mass on the green or yellow band at $\eta>1$ by adjusting $\kappa$ or $\lambda$ to be smaller than the value used in (b) of Fig. \ref{fig:k=1d0 and k=15.6d6} to get a brane with de Sitter background.}

It is shown in Ref.~\cite{Delsate:2012ky} that under certain assumptions, it is possible to re-express the energy momentum tensor modified by the EiBI theory in the form of perfect fluid also but in terms of  $q_{\mu\nu}$. In this form the pressure and the energy density become apparent pressure \textcolor{black}{$p_q$} and energy density $\epsilon_q$. In this view, EiBI can be considered as GR with additional isotropic gravitational pressure $\mathcal{P}$ in apparent stress tensor {$\bar{T}^\mu_\nu$}. In the following discussions, we use the same procedure as one used in Ref.~\cite{Delsate:2012ky} to obtain the brane correction of  {$p_q$} and $\epsilon_q$. We then calculate the energy conditions from Ref. \cite{Delsate:2012ky} to study the allowed range for $w,\lambda,\kappa,$ and $\eta$. In this model the {$p_q$} and $\epsilon_q$ become 
\begin{eqnarray}
&&\text{"apparent energy density"}=\epsilon_\text{eff}, \text{~and~} \nonumber\\
&&\text{"apparent pressure"}=p_\text{eff}+{4\mathcal{P}\over (8\pi G_N c^{-2})^2\lambda}.
\end{eqnarray}
We need to note that in Eq. \eqref{eq:0001} $p_\text{eff}+4\mathcal{P}/(8\pi G_N c^{-2})^2\lambda$ affects $p'(r)$ and in Eq. \eqref{eq:0002} $\epsilon_\text{eff}$ affects $m$. The following energy conditions will be used to study the allowed range for $w,\lambda,\kappa,\eta$:
\begin{enumerate}
\item null energy condition (NEC): $$\epsilon_\text{eff}+\left(p_\text{eff}+{4\mathcal{P}\over(8\pi G_N c^{-2})^2\lambda}\right)\geq0,$$
\item weak energy condition (WEC): NEC and $\epsilon_\text{eff}\geq0$,
\item strong energy condition (SEC): NEC and $$\epsilon_\text{eff}+3\left(p_\text{eff}+{4\mathcal{P}\over(8\pi G_N c^{-2})^2\lambda}\right)\geq0,$$
\item dominant energy condition (DEC): $$\epsilon_\text{eff}-\left|p_\text{eff}+{4\mathcal{P}\over(8\pi G_N c^{-2})^2\lambda}\right|\geq0,\text{ and}$$
\item causal energy condition (CEC): $$|\epsilon_\text{eff}|-\left|p_\text{eff}+{4\mathcal{P}\over(8\pi G_N c^{-2})^2\lambda}\right|\geq0.$$
\end{enumerate}

The reason why we consider these energy conditions in terms of the ``apparent", rather than the physical, energy and pressure is that we wish to establish the upper bound of $\kappa$ purely from within the theory itself. In order for the EiBI theory to be consistent it should reduce to GR in the weak-coupling limit, $\kappa\rightarrow0$. On the other hand, there should be an upper limit for $\kappa$ so that the success of GR in astrophysics is not spoiled. So far such upper bound comes from observation of compact star, $\kappa<1.4\times10^8\ \text{m}^2$~\cite{Avelino12}. This is quite a large value that may not immediately be verified or ruled out in the near future with our current technology. Our idea is the following. Instead of looking for constraint from astrophysics, we look for constraint from within the internal consistency of the theory itself. It is well-known in EiBI theory~\cite{Berti_etal2015} that there is a degeneracy in the field equations between 
\begin{equation}
\label{twin}
\left(G^{\mu\nu}\right)_{EiBI}={8\pi G_N\over c^2}\kappa \left(T^{\mu\nu}\right)_{physical} \ \ \ \leftrightarrow\ \ \ \left(\bar{G}^{\mu\nu}\right)_{GR}={8\pi G_N\over c^2}\left(\bar{T}^{\mu\nu}\right)_{apparent},
\end{equation}
where the left-side equation comes from~\eqref{eq:metr},
\begin{equation}
\left(G^{\mu\nu}\right)_{EiBI}\equiv\eta g^{\mu\nu}-{1\over\tau}\left(g_{\mu\nu}+\kappa\bar{R}_{\mu\nu}\right)^{-1},
\end{equation}
 and the right-side one comes from~\eqref{degen}. The physical observables extracted from both equations are indistinguishable. While we do not (yet) have any established constraint for the left-side equation in~\eqref{twin}, we know that such standard constraint exists for GR. Thus, by constraining the (apparent) energy-momentum tensor by means of positivity of energy conditions for the right-side equation if~\eqref{twin}, we hope to extract an upper bound for $\kappa$.

\begin{figure}[t]
\centering
\includegraphics[width=.45\linewidth]{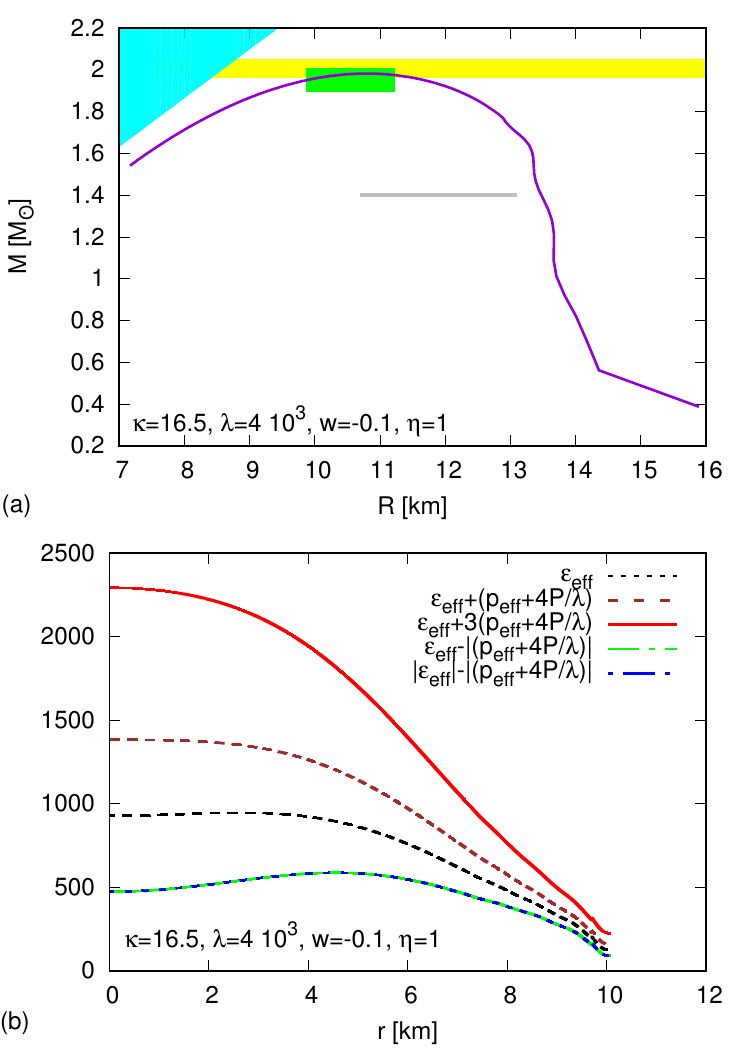}
\caption{Mass-radius relation (a) which is inside the regions from two observations and its plot on energy conditions (b). These conditions are satisfied. (Here {$\mathrm{P}=\mathcal{P}/(8\pi G_N c^{-2})^2$} and $\eta=1$.)}
\label{fig:observed}
\end{figure}

\begin{figure}[t]
\centering
\includegraphics[width=.45\linewidth]{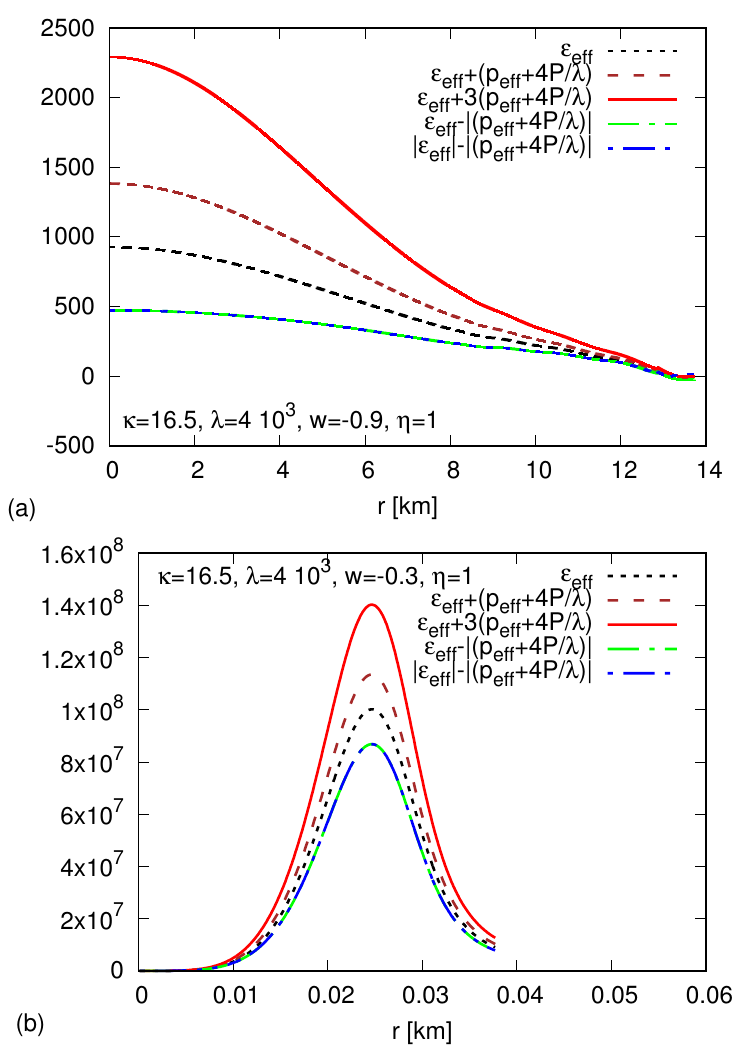}
\caption{Profiles from $p_c=200$  MeV $\rm fm^{-3}$ with different $w$. (a) violates all energy conditions but (b) does not, yet at $-0.5<w\leq-0.3$ gives a relatively radius that is too small. Hence we obtain $w<-0.9$ or $w>-0.3$. Figure (a) is the case when either NEC, WEC, SEC, DEC, or CEC are violated by a certain value of $w$.}
\label{fig:analysis_w}
\end{figure}

\begin{figure}[t]
\centering
\includegraphics[width=.45\linewidth]{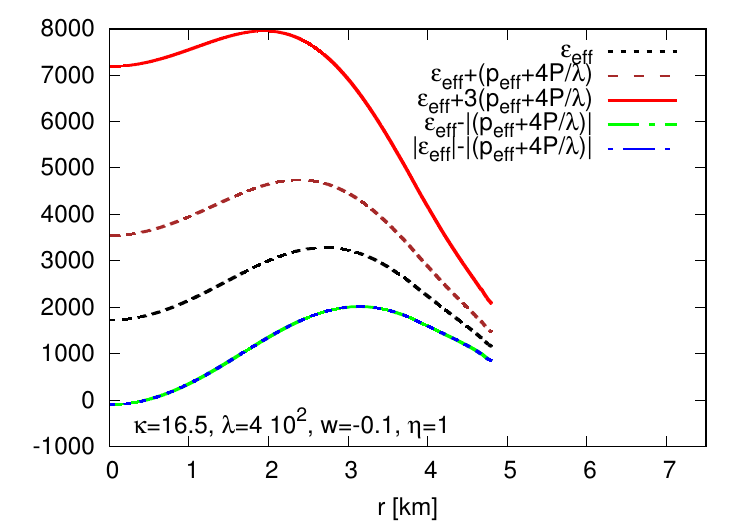}
\caption{A profile from $p_c=200$  MeV $\rm fm^{-3}$ with different $\lambda$ that violates at least one of the energy conditions. We obtain that at $\lambda=4\times10^2$, it violate DEC and CEC thus $\lambda>4\times10^2$. The figure is the case when either NEC, WEC, SEC, DEC, or CEC are violated by a certain value of $\lambda$.}
\label{fig:analysis_lambda}
\end{figure}

\begin{figure}[t]
\centering
\includegraphics[width=.45\linewidth]{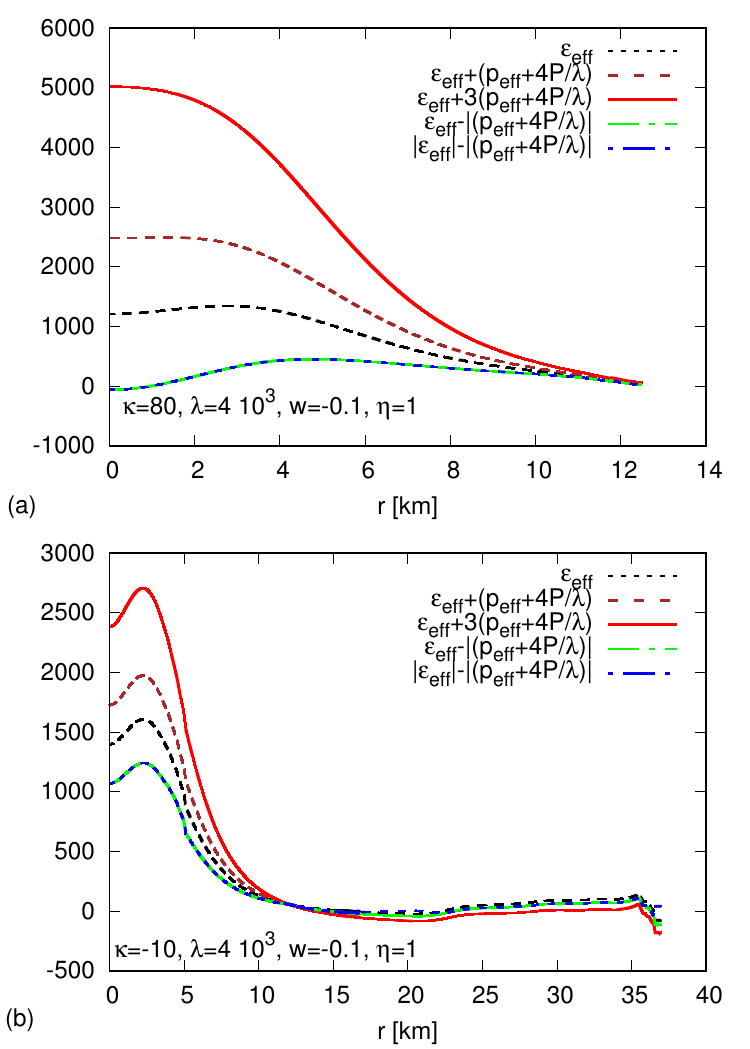}
\caption{Profiles from $p_c=200$ MeV $\rm fm^{-3}$ with different $\kappa$. (a) imply violation of DEC and CEC while from (b) implies violation of SEC, DEC, and CEC. Hence we obtain that $-10<\kappa<80$. These figures are the case when either NEC, WEC, SEC, DEC, or CEC are violated by a certain value of $\kappa$.}
\label{fig:analysis_kappa}
\end{figure}

\begin{figure}[t]
\centering
\includegraphics[width=.45\linewidth]{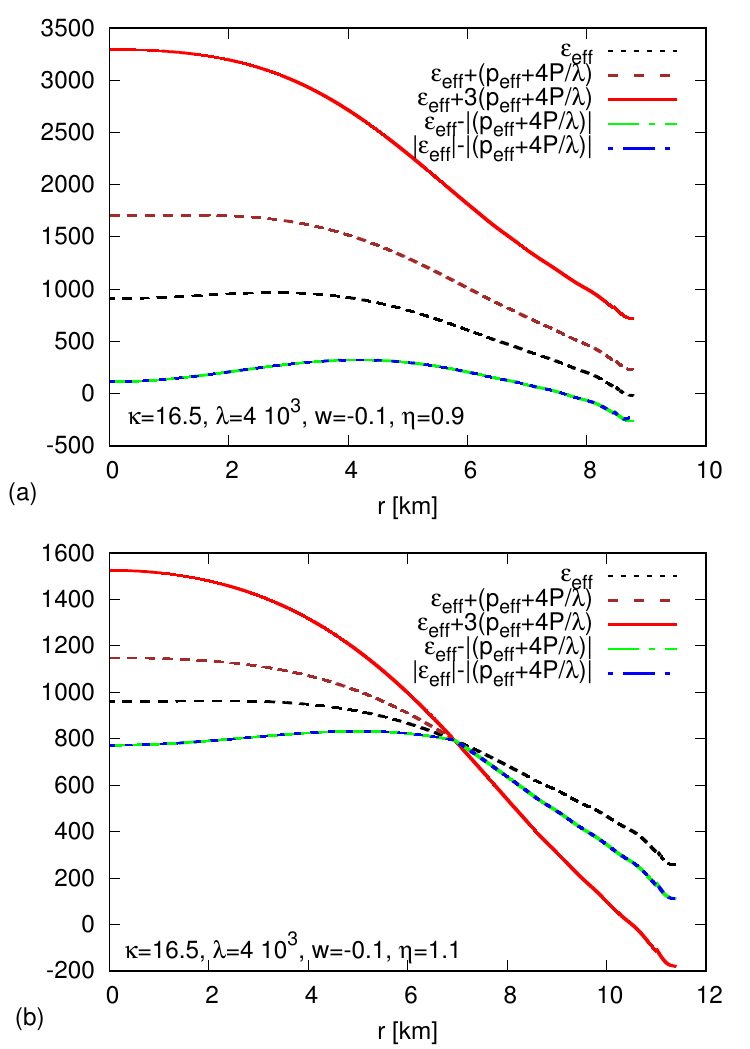}
\caption{Two profiles from $p_c=200$  MeV $\rm fm^{-3}$ with different $\eta$ and we obtain $0.9<\eta<1.1$, whose lower and upper bound is from (a) and (b) respectively.These figures are the case when either NEC, WEC, SEC, DEC, or CEC are violated by a certain value of $\eta$.}
\label{fig:analysis_eta}
\end{figure}

The example and analysis results are shown in Fig.~\ref{fig:observed} and Figs.~\ref{fig:analysis_w}-\ref{fig:analysis_eta}, respectively. Note that in the corresponding figures,  the forbidden region is below the horizontal axis (vertical axis = 0). If a part of any curve is in this region, then the energy conditions are violated.

In Fig. \ref{fig:observed}, we show an example of mass-radius relation in (a) whose energy conditions in (b) are not violated from setting the parameter to a certain value: $w=-0.1$, $\lambda=4\times10^3$, $\kappa=16.5$, and $\eta=1$ and using {$p_c=200$ MeV $\rm fm^{-3}$}. In Figs. \ref{fig:analysis_w}-\ref{fig:analysis_eta}, in principle  we vary the value of one of these constants while the others are kept fixed. By this method we obtain the allowed region from each $w,\lambda,\kappa$ and $\eta$. The results can be summarized in the following lists.

\begin{enumerate}
\item
We vary $w$ whose results are shown in Fig. \ref{fig:analysis_w}. It can be seen that $w=-0.5$ makes $\mathcal{U}'(r)$ become singular and $w\rightarrow\pm\infty$ makes $\mathcal{U}$ constant. Hence we only need to check around $w=-0.5$. The allowed value from the figure is $w<-0.9$ or $w>-0.5$, but since at $-0.5<w\leq-0.3$ gives very small radius, we then remove this region and obtain that $w<-0.9$ or $w>-0.3$ which in turn justifies why the authors of \cite{Castro:2014xza} choose $w=-3,-1,-0.6,-0.1,0.2,2$ and exclude the region around $w$=-0.5 in calculating mass-radius curves.

\item
In Fig. \ref{fig:analysis_lambda} we vary $\lambda$. We just need a lower bound for $\lambda$ since the bulk contributions vanishes when $\lambda\rightarrow\infty$. From Fig. \ref{fig:analysis_lambda}, we can seen that $\lambda>4\times10^{2}\ \text{MeV}/\text{fm}^3=(4/5.07^3)\times10^{11}\ \text{MeV}^4\simeq 3.07\times10^{9}\ \text{MeV}^4$ obeys the energy conditions, and this lower bound is much larger than that found in Ref. \cite{Garcia-Aspeitia:2013jea}.

\item
In Fig. \ref{fig:analysis_kappa} we vary $\kappa$. This EiBI's nonlinearity constant is firstly thought to be positive valued and thus has an upper bound. While this is true, $\kappa$ turns out can also be negative valued and still satisfy the above energy conditions until a certain large negative value. In the figures, we obtain that $-10<\kappa<80$ still obey energy conditions. It still contains range of $0<\kappa<6$ allowed by astrophysical-cosmological constraints in Ref. \cite{Avelino12}.

\item
We also vary $\eta$ as shown in Fig. \ref{fig:analysis_eta}. Recall that here we vary by $\eta=1\pm\delta\eta$ with $\delta\eta>0$. At $\delta\eta=0.01$, we have found that no violation on the energy conditions, but at $|\delta\eta|=0.1$ both figures show at least one violation. Hence we obtain that $0.9<\eta<1.1$ obey the energy conditions.

\end{enumerate}

\begin{figure}[t]
\centering
\includegraphics[width=.45\linewidth]{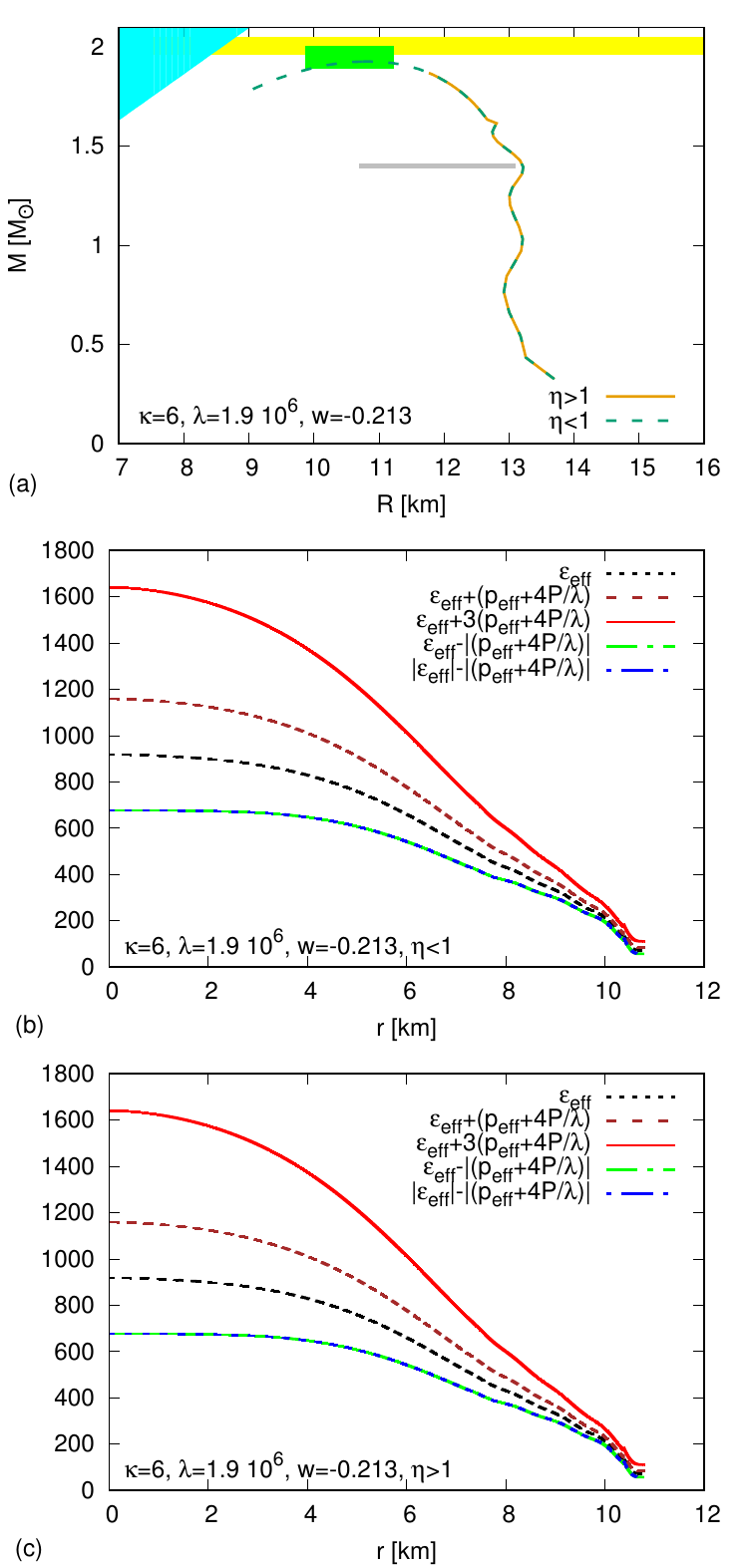}
\caption{Mass-radius curve (a) which is inside the green-shaded regions with $\eta=1+\kappa H$ and $H=\pm2.08\times10^{-46}~[10^{-6} \text{m}^{-2}]$. Both energy conditions plots from a profile with $p_c=200$ MeV $\rm fm^{-3}$ (b) and (c), respectively, for anti-de Sitter and de Sitter brane.}
\label{fig:cosmo}
\end{figure}

From here one could also investigate the impact of $\eta$ by setting $H$ at the observed value of cosmological constant in Refs. \cite{Carroll:2000fy,Carroll:2004st}
\begin{equation}
\rho_H={H\over 8\pi G_N}\sim10^{-8} {\text{erg}\over\text{cm}^3}.
\label{eq:cosmo}
\end{equation}
This implies $
H\sim2.08\times10^{-46}~[10^{-6} \text{m}^{-2}],$ which then gives the value of $\eta$ by $\eta=1+\kappa H>1$.
By setting $\kappa=6$ (the allowed value of $\kappa$ by observation is $0<\kappa\leq6$ \cite{QISR2016}), we increase $\lambda$ to increase the mass and decrease $w$ to decrease the radius to obtain the maximum mass in the observed value denoted by the green-shaded region. The mass-radius result and its profile at $p_c=200$ MeV $\rm fm^{-3}$ is shown in Fig. \ref{fig:cosmo}(a)($\eta>0$) from $H=2.08\times10^{-46}~[10^{-6} \text{m}^{-2}]$. 
Fig. \ref{fig:observed}(b) imply that there is no violation on any of the energy conditions. {Here we observe that by setting the tension $\lambda$ to be one order higher than $10^3$ MeV/fm$^3$, i.e. $\lambda\sim10^4$ MeV/fm$^3$, we need $w$ to approach $w\rightarrow-0.5$ from $w=-0.1$ to decrease the radius. The wiggle that appears at some small $p_c$ in the radius-mass relation at $w=-0.1$ is evident and it becomes more wiggly as $w<-0.1$. This wiggle is independent of $\lambda$, i.e., it just increases both the maximum mass and its radius. Note that in Fig. \ref{fig:observed} $w=-0.1,~\lambda\sim10^3$ is used while in Fig. \ref{fig:cosmo} $w=-0.213,~\lambda\sim10^6$. 
Interestingly, the mass-radius curve fairly compatible with three observational constraints of $\lambda$ \cite{Steiner:2010fz,Ozel:2016oaf,Antoniadis:2013pzd} used.} 
We also show the mass-radius relation brane with anti-de Sitter background case in Fig. \ref{fig:cosmo}(a) ($\eta<1$) from $H=-2.08\times10^{-46}~[10^{-6} \text{m}^{-2}]$ and this case is also allowed by the energy conditions constraints in Fig. \ref{fig:cosmo}(c). In contrast to Fig. \ref{fig:eta_neq_1}, the two curve in Fig. \ref{fig:cosmo}(a) are not separated since $\kappa H \sim 10^{-46}$ making $\eta$ still very near to $1$.\\

\section{Conclusions}
\label{sec_conclu}

In this work, we combine two new physics proposals, the braneworld and the modified gravity (EiBI) theories, to study the inner structure of NS. Our work is an extension of the previous result~\cite{QISR2016} by extending it to higher dimension. The bulk is assumed to be empty, while the matter fields live on the brane. The field equations can be cast into the usual Einstein's GR with ``effective" energy-momentum tensor. The EoS at the core and at the crust are calculated using the same prescription as in~\cite{QISR2016}.  

It is known from~\cite{Castro:2014xza} that the existence of higher-dimensional bulk reduces the NS compactness. From the point of view of astrophysical signatures this is depressing, since most observational results seem to favor $M\gtrsim2.1 \text{M}_\odot$. This is precisely our motivation to invoke the EiBI theory in braneworld. In our model, the compatibility with observational data can be restored by adjusting $\kappa$. We found that $\kappa$ and $\lambda$ work in sort of the opposite way. By balancing each other we can have mass and radius of NS compatible with observations still within the cosmologically- and astrophysically-accepted range, $0<\kappa<6\times 10^6\ \text{m}^2$ and $\lambda\gg1\ \text{MeV}^4$.

As with other TOV-braneworld proposals, our model is dependent on $w$ that comes from the ``Weyl EoS'' $\mathcal{P}=w\mathcal{U}$ taken from~\cite{Castro:2014xza,Felipe:2016lvp}. By setting $w=0$, the ``apparent'' energy-momentum tensor ${T^m_n}_\text{eff}$ is isotropic.  Here we also investigate the model when $w\neq0$, making ${T^m_n}_\text{eff}$ anisotropic. It was found in Ref.~\cite{Castro:2014xza} that $w$ increase or decrease the radius of the star depending on its value. We invoked the energy conditions from Ref. \cite{Delsate:2012ky} to determine the allowed range of values of $\kappa,\lambda,$ and $w$. We obtain that $-10\times 10^6~\text{m}^2<\kappa<80\times 10^6~\text{m}^2$, $\lambda>4\times10^{2}\ \text{MeV}/\text{fm}^3\simeq 3.07\times10^9\ \text{MeV}^4$, $w<-0.9$ or $w>-0.3$, and $0.9<\eta<1.1$.

We also need to make some comments on the curvature of the bulk. Setting the projection of bulk's cosmological constant onto the brane to be zero, $\bar{\Lambda}=0$, we necessarily have anti-de Sitter bulk, ${\Lambda} =- {2^{7}\cdot 3\pi^2 G_N^2/\kappa_5^8 c^4}$. This is nothing but the fine-tuning problem that plagues the Randall-Sundrum scenario. However, we can still set the brane to be de Sitter or anti-de Sitter by tweaking $\eta$. We found that, by setting $\eta>1$ ($\eta<1$), the mass-radius relation is also lifted (lowered). Now, the whole results above might change should we set instead $\eta=0$ and $\bar{\Lambda}\neq0$. This enables the bulk to be other than AdS at the cost of the curvature of the brane. The $4d$ cosmological effect is the same (that the brane feels some effective cosmological constant), but the astrophysical signatures ({\it e.g,} NS mass-relation) might be different since the appearance of $\bar{\Lambda}$ changes the TOV equations. This possibility deserves further investigation.\\

\appendix

\section{Eddington inspired Born-Infeld (EiBI) gravity}
\label{eibi} 

Here we discuss a gravity-modified model where the scalar curvature is dependent on the connection. Let us consider a Born-Infeld type of Lagrangian density for Ricci tensor with \cite{Banados:2010ix}
\begin{eqnarray}
\mathcal{S}={c^3\over 16\pi G_N}{2\over \kappa}\int d^4 x \left[\sqrt{-\det(g_{\mu\nu}+\kappa \bar{R}_{\mu\nu})}\right.
-\left.\eta\sqrt{-\det(g_{\mu\nu})}\right]+\mathcal{S}_m.\label{eq:LagEiBI}
\end{eqnarray}
$\kappa$ and $\eta$ are parameters corresponding to nonlinearity and cosmological constant, respectively. The last term is from matter contribution. At $\kappa\rightarrow0$, one obtain the Einstein-Hilbert action
\[
\mathcal{S}={c^3\over 16\pi G_N}\int d^4 x \left[ \bar{R}-2H\right]\sqrt{-\det(g_{\mu\nu})}+\mathcal{S}_m,
\]
with cosmological constant $H={(\eta-1)/\kappa}.$ It is also easy to show that when $\mathcal{S}_m=0$ the Action~\eqref{eq:LagEiBI} reduces to the ordinary Einstein-Hilbert's (for example, see~\cite{Banados:2010ix, Delsate:2012ky}.) Since the matter fields are assumed to live only on the brane, the gravity on the fifth dimension is identical to GR. On the brane, on the other hand, we have the full EiBI gravity. 

Here we invoke the Palatini formalism , i.e., connection and metric are defined as separate entities. As explained in~\cite{Banados:2010ix,Sotani14,Vollick:2003qp,Dadhich:2000am}, the Palatini formalism is employed to avoid the appearance of fourth-order field equations with ghost terms should we use the usual metric variation~\cite{Deser:1998rj}. In this formalism, the connection $\Gamma^{\alpha}_{\mu\nu}$ is dependent on an ``auxilliary" metric $q_{\mu\nu}$. The Ricci tensor still depends on the connection. The matter, on the other hand, couples to the physical metric $g_{\mu\nu}$. Varying the action~\eqref{eq:LagEiBI} with respect to the metric $g^{\mu\nu}$ and connection $\Gamma^\alpha_{~\mu\nu}$ respectively, we have~\footnote{To vary the action with respect to $\Gamma^\alpha_{~\mu\nu}$, one can write the connection as $\Gamma^\alpha_{~\mu\nu}=\tilde{\Gamma}^\alpha_{~\mu\nu}+C^\alpha_{~\mu\nu},$ which is symmetric (torsion-free), with $\tilde{\Gamma}^\alpha_{~\mu\nu}$ and $C^\alpha_{~\mu\nu}$ a connection and a tensor, respectively. After varying, $C^\alpha_{~\mu\nu}$ vanishes and $\tilde{\Gamma}^\alpha_{~\mu\nu}$ must be metric compatible~\cite{Carroll:2004st}.}
\begin{equation}
(g_{\mu\nu}+\kappa \bar{R}_{\mu\nu})^{-1}
={\tau} \left(\eta g^{\mu\nu} -{8\pi G_N\over c^2}\kappa T^{\mu\nu}\right)\label{eq:metr}
\end{equation}
\begin{equation}
q_{\mu\nu}=g_{\mu\nu}+\kappa \bar{R}_{\mu\nu},\label{eq:relationmetric}
\end{equation}
where $\bar{R}_{\mu\nu}$ depends on the connection
\begin{equation}
\Gamma^{\alpha}_{~\mu\nu}={1\over2}q^{\alpha\kappa}(q_{\nu\kappa,\mu}+q_{\kappa\mu,\nu}-q_{\mu\nu,\kappa}).
\end{equation}
Here
\begin{align}
T^{\mu\nu}\equiv&{2\over\sqrt{-\det g_{\mu\nu}}}{\delta\mathcal{S}_m\over \delta g_{\mu\nu}},\label{eq:physicalT}\\
\tau\equiv&{\sqrt{-\det(g_{\mu\nu})}\over \sqrt{-\det(g_{\mu\nu}+\kappa \bar{R}_{\mu\nu})}}.
\end{align}

We can reproduce the Einstein's equation using the expression of Ricci tensor
\begin{equation}
\bar{R}^\mu_\nu= {(1-\tau\eta)\over \kappa} \delta^\mu_\nu + {8\pi G_N\over c^2}\tau T^\mu_\nu, \label{eq:ein}
\end{equation}
which makes
\begin{equation}
\bar{G}^\mu_\nu= {8\pi G_N\over c^2}\tau T^\mu_\nu-\left({8\pi G_N\over c^2}\tau {T\over2}+{(1-\tau\eta)\over \kappa} \right) \delta^\mu_\nu,\label{eq:ein2}
\end{equation}
with $T^\mu_\nu=T_{\kappa\nu}g^{\mu\kappa},~T=T_{\mu\nu}g^{\mu\nu}$ and
\begin{equation}
\tau={1\over \sqrt{\det\left(\eta \delta^\mu_\nu-{8\pi G_N\over c^2}\kappa T^\mu_\nu\right)}}.
\end{equation}
This becomes the Einstein equation when $\kappa\rightarrow 0$ and, simultaneously, $\eta\rightarrow1$.
For convenience, we also define
\begin{equation}
\bar{T}^\mu_\nu= \tau T^\mu_\nu-\left(\tau {T\over2}+{(1-\tau\eta)\over (8\pi G_N c^{-2})\kappa} \right) \delta^\mu_\nu, 
\end{equation}
so that \eqref{eq:ein2} become
\begin{equation}
\label{degen}
\bar{G}^\mu_\nu={8\pi G_N\over c^2}\bar{T}^\mu_\nu,
\end{equation}
which is interpreted as Einstein equation viewed from metric $q_{\mu\nu}$. 
Notice that $\bar{T}_{\mu\nu}$ is conserved under covariant derivative of metric $q_{\mu\nu}$. This is different from $T_{\mu\nu}$ from \eqref{eq:physicalT} which is conserved under covariant derivative of metric $g_{\mu\nu}$.

Now let us define, respectively, the physical and auxiliary metric to be homogeneous and isotropic as
\begin{eqnarray}
g_{\mu\nu}dx^\mu dx^\nu&=&-A^2(r)c^2dt^2+B^2(r)dr^2+C^2(r)d\Omega^2_2,~~~~~~~~\\
q_{\mu\nu}dx^\mu dx^\nu&=&-F^2(r)c^2dt^2+G^2(r)dr^2+r^2d\Omega^2_2.\label{eq:metricq}
\end{eqnarray}
Assuming the matter in manifold with physical metric is an ideal fluid
\begin{equation}
T^{\mu\nu}=(\rho+p c^{-2})u^\mu u^\nu+pc^{-2}g^{\mu\nu},
\end{equation}
where $\rho$ and $p$ the physical energy density and pressure which are positive semi-definite, respectively, with velocity vector $u^\mu=(\sqrt{-g^{00}},0,0,0),~u^\mu u_\mu=g_{\mu\nu}u^\mu u^\nu=-1$ thus we obtain 
\begin{equation}
\tau={1\over ab^3},~~ a=\sqrt{\eta+8\pi G_Nc^{-2}\kappa\rho},~~ b=\sqrt{\eta-8\pi G_Nc^{-4}\kappa p}.
\label{eq:ab}
\end{equation}
Its nonzero components are
\begin{eqnarray}
\bar{T}^0_0&=&{-a^2+3b^2-2ab^3 \over2ab^3 (8\pi G_N c^{-2})}\equiv -\rho_q, \label{eq:edenapp}\\
\bar{T}^r_r=\bar{T}^\theta_\theta=\bar{T}^\phi_\phi&=&{ a^2+b^2-2ab^3\over2ab^3 (8\pi G_N c^{-2})}\equiv p_q c^{-2}, \label{eq:pressapp}
\end{eqnarray}
with $\rho_q$ and $p_q$ are the apparent energy density and pressure, respectively. Both apparent energy density and apparent pressure are no longer positive semi-definite.
The Ricci tensor nonzero components from the auxiliary metric are
\begin{eqnarray}
\bar{R}^0_0&=&-G^{-2}\left[{F''\over F}-{F'G'\over FG} +2{F'\over Fr}\right],\\
\bar{R}^r_r&=&-G^{-2}\left[{F''\over F}-{F'G'\over FG} -2 {G'\over Gr}\right],\\
\bar{R}^\theta_\theta&=&-G^{-2}\left[
{1\over r}\left\{
{F'\over F}-{G'\over G}\right\}+{1\over r^2} \right]+{1\over r^2}.
\end{eqnarray}
From \eqref{eq:ein} we have
\begin{eqnarray}
{(1-\tau\eta)\over \kappa} -(8\pi G_N c^{-2})\tau \rho&=&\bar{R}^0_0,\label{eq:001}\\
{(1-\tau\eta)\over \kappa} +(8\pi G_N c^{-2})\tau pc^{-2}&=&\bar{R}^r_r,\label{eq:002}\\
{(1-\tau\eta)\over \kappa} +(8\pi G_N c^{-2})\tau pc^{-2}&=&\bar{R}^\theta_\theta,\label{eq:003}
\end{eqnarray}
and from using an ansatz
\begin{equation}
G^{-2}=1-{2G_N c^{-2}m(r)\over r}
\end{equation}
we obtain
\begin{equation}
{F'\over F}={-2r^3+r^3(a^2+b^2)/ab^3+4\kappa G_N c^{-2}m \over 4\kappa r (r-2G_N c^{-2}m)}
\label{eq:metric01}
\end{equation}
and
\begin{equation}
m'(r)= {r^2\over 2\kappa G_N c^{-2}}\left[ 1 +{a^2-3b^2\over 2ab^3}\right].
\end{equation}
To obtain TOV equation, we need $\nabla_b \bar{G}^{ab}=0$ which if we set $a=r$ we obtain $(\bar{G}^r_r)'+{F'\over F}(\bar{G}^r_r-\bar{G}^0_0)=0,$
or
\begin{equation}
p'(r)
=-{F'\over F}{b\over 2\pi G_N c^{-4} \kappa} \left[
{ab(a^2-b^2) \over {4ab^2+(3a-b{da\over db})(a^2-b^2)}}  
\right].
\end{equation}

We can also use conservation on energy-momentum tensor and relation between metric \cite{QISR2016,Harko13} to obtain their expression.
The relation between the two metric can be seen from Eq. \eqref{eq:metr} or explicitly
\begin{equation}
A^2=ab^{-3}F^2,~
B^2=G^2/ab,~
C^2=r^2/ab.
\label{eq:relation01}
\end{equation}
Be aware that $T^{ab}$ is on manifold with metric $g_{ab}$ not $q_{ab}$ thus by $\nabla_a T^{ab}=0$ we obtain $p'c^{-2}=-(\rho+p c^{-2})A'/A$ or
\begin{equation}
{A'\over A}={2b'b\over a^2-b^2}.
\end{equation}
By differentiating the first equation in \eqref{eq:relation01} with respect to $r$ multiplying it by $A^2$ we have
\begin{equation}
{2F'\over F}={2A'\over A} - {a' \over a} + {3b' \over b}.
\end{equation}
Defining speed of sound as
\begin{equation}
c^2_q={da\over db}=-{b\over a}{d\rho\over d(pc^{-2})},
\end{equation}
we obtain
\begin{eqnarray}
{2F'\over F}=\left[{4b\over a^2-b^2} - {c_q^2\over a} + {3\over b}\right]b'\nonumber\\
=\left[{4ab^2+(3a-c_q^2 b)(a^2-b^2)\over ab(a^2-b^2)}\right]b'.
\end{eqnarray}
Substituting $F'/F$ with \eqref{eq:metric01} we obtain
\begin{eqnarray}
p'(r)={-b\over 4\pi G_Nc^{-4} \kappa} {[ab(a^2-b^2)]\over[4ab^2+(3a-c_q^2 b)(a^2-b^2)]}\nonumber\\
{[r^3(-2+a/b^3+1/ab)/ 2\kappa+2 G_N c^{-2}m] \over r (r-2G_N c^{-2}m)}.~~~
\end{eqnarray}
Numerical computation then can be done with a choice of equation of state $p=p(\rho)$.

\acknowledgments

HSR thanks Muhammad Iqbal and Reyhan Lambaga for the discussions on EiBI theory. This work is partially supported by Universitas Indonesia.


\end{document}